\documentclass[aps,pra,twocolumn,showpacs,superscriptaddress,groupedaddre
ss]{revtex4-1}
\pdfoutput=1

\usepackage{braket,amsmath,amsfonts,hyperref,graphicx,subcaption}
\newcommand{\tr}{\mathrm{tr}}
\newcommand{\parder}[2]{\frac{\partial {#1}}{\partial {#2}}}
\newcommand{\pa}{\partial}
\newcommand{\ot}{\otimes}
\newcommand{\id}{\mathbb{I}}
\newcommand{\n}{\noindent}
\newcommand{\al}{\alpha}\newcommand{\be}{\beta}
\newcommand{\la}{\lambda}
\newcommand{\si}{\sigma}\newcommand{\om}
{\omega}

\begin{document}

\title{Universal entanglement timescale for R\'enyi entropies}

\author{Jesse C. Cresswell}
\email{jcresswe@physics.utoronto.ca}

\affiliation{Department of Physics, University of Toronto, Toronto, ON, M5S 
1A7, Canada}

\date{\today}

\begin{abstract}
Recently it was shown that the growth of entanglement in an initially 
separable state, as measured by the purity of subsystems, can be 
characterized by a timescale that takes a universal form for any Hamiltonian. 
We show that the same timescale governs the growth of 
entanglement for all R\'enyi entropies. Since the family of R\'enyi entropies 
completely characterizes the entanglement of a pure bipartite state, our 
timescale is a universal feature of bipartite entanglement. The timescale 
depends only on the interaction Hamiltonian and the initial state.
\end{abstract}


\maketitle

\section{Introduction}

\normalsize

Composite quantum systems exhibit correlations among subsystems which 
cannot be explained in terms of classical probabilities. For pure states, these 
quantum correlations are known as entanglement. In this paper, we study how 
entanglement is generated by the mutual interactions among 
subsystems as the overall state evolves in time. 

The time evolution of entanglement has become a focus in a variety of 
research fields. Its early study in quantum optical systems 
\cite{Phoenix1988,Gea-Banacloche1990} has bloomed into a major area of 
research in many-body and condensed-matter systems 
\cite{Calabrese2004,Calabrese2009,Casini2009,Amico2008,
Laflorencie2015}, 
and conformal field theories dual to theories of quantum gravity 
\cite{Maldacena1999,Ryu2006,Hubeny2007}. For some classes of systems, 
general features have been found, including scaling laws 
\cite{Eisert2006,Bravyi2006} and generic linear growth 
\cite{Calabrese2005,Liu2014a,Hartman2013,Bianchi2017}.

The growth of entanglement is especially important in experimental systems 
where entanglement between the system and its environment leads to 
decoherence \cite{Zurek2003}. A complete understanding of the evolution of 
entanglement requires solving the dynamics of the overall state. This is often 
not feasible, including for decoherence where the Hamiltonian describing 
interactions with the environment is not known explicitly.

It is therefore interesting to ask what aspects of entanglement growth, if any, 
are shared by all quantum systems. Broad statements can be made 
in this direction with minimal assumptions about system 
dynamics by relying on special initial conditions instead.

To begin, bipartite entanglement between subsystems must be defined 
with respect to a partition of the system's degrees of 
freedom, represented as a fixed factorization of the Hilbert 
space $\mathcal{H}=\mathcal{H}_A\ot \mathcal{H}_B$. The 
Hamiltonian for the full system can be expressed as
\begin{equation}\label{eqHam}
H=\sum_n A_n\ot B_n,
\end{equation}
where each $A_n$ is an operator acting on subsystem $\mathcal{H}_A$, and each 
$B_n$ acts on $\mathcal{H}_B$. Any number of terms may be 
included as long as $H$ is Hermitian. Since the algebra of 
operators acting on $\mathcal{H}$ is isomorphic to the tensor product of 
subsystem algebras, any Hamiltonian can be 
represented this way \cite{Zanardi2001}.

Recently it was shown by Yang \cite{Yang2017a} that starting from a 
pure, unentangled state
\begin{equation}\label{eqSt}
\ket{\Psi(0)}=\ket{\psi(0)}_A\ot\ket{\psi(0)}_B,
\end{equation}
 the growth of entanglement under the unitary evolution generated by 
 \eqref{eqHam} is characterized by a universal timescale,
 \small
\begin{equation}\label{eqTS}
T_{\mathrm{ent}}{=}
\left[\sum_{n,m}\left(\langle A_nA_m \rangle{-}\langle A_n \rangle\langle A_m 
\rangle\right)\left(\langle B_nB_m \rangle{-}\langle B_n \rangle\langle B_m 
\rangle\right)\right]^{-\tfrac{1}{2}}.
\end{equation}
\normalsize
Here the expectation values are taken in the initial state. The timescale is 
universal in the sense that it takes this form for any quantum system that 
satisfies the requirements \eqref{eqHam} and \eqref{eqSt}. The 
entanglement timescale was derived by studying one particular measure of 
the entanglement between subsystems $A$ and $B$, namely, the purity 
$P(\rho_A)=\tr_A \rho_A^2$ of the reduced density matrix $\rho_A=\tr_B 
\rho$. By the assumption \eqref{eqSt}, the purity is initially maximal so that its 
dynamics are governed at lowest order in $t$ by $d^2P/dt^2$. The second 
derivative is proportional to $T_{\mathrm{ent}}^{-2}$ which is entirely 
determined by the expectation values of the interaction Hamiltonian operators 
in the initial state. 

In this paper, we show that the same entanglement timescale \eqref{eqTS} 
governs the growth of entanglement as measured by the entire family of 
quantum R\'enyi entropies \cite{Renyi1961},
\begin{equation}\label{eqRenyi}
S_\al(\rho_A)=\frac{1}{1-\al} \ln \tr_A \rho_A^\al,
\end{equation}
 where $\al$ is taken to be a positive integer. As a family, the R\'enyi entropies 
provide complete information about the eigenvalue distribution of the reduced 
density matrix $\rho_A$, and hence completely characterize the 
entanglement in an overall pure, bipartite state \cite{Li2008,Headrick2010}. 
Therefore, the entanglement timescale \eqref{eqTS} is a universal feature of 
bipartite entanglement. 

The most common measure of entanglement, the entanglement entropy 
{$S(\rho_A)\ {=}-\tr_A( \rho_A\ln \rho_A)$}, corresponds to the $\al\to 1$ limit of 
\eqref{eqRenyi}. Its second time derivative can be obtained by an analytic 
continuation in $\al$ from our general results for $\al\geq2$ after which 
\eqref{eqTS} appears with a logarithmically divergent prefactor, reflecting the 
sensitivity of $S(\rho_A)$ to small eigenvalues of the density matrix. We 
provide an example of these results by working with the Jaynes-Cummings 
model \cite{Shore1993}.

Notably, the entanglement timescale can be computed without the need to 
solve for the dynamics of the system. For a given experimental preparation of 
an unentangled state, our results provide an easily calculable estimate of 
when entanglement will become significant. Advances in the optical control of 
atoms have led to the first direct measurement of a R\'enyi entropy in a 
many-body system, and subsequently to measurements of its growth 
\cite{Daley2012,Schachenmayer2013,Islam2015a,Kaufman2016,Elben2018}. 
We return to these measurements for comparison to the entanglement 
timescale in Sec. \ref{dis}.

\section{The entanglement timescale for R\'enyi entropies}\label{sec2}

To begin, we briefly review the relevant properties of $S_\al(\rho_A)$ defined 
in \eqref{eqRenyi}. For any positive integer $\al$, $S_\al(\rho_A)$ is an 
entanglement measure that is minimized at zero if and only if the total state $
\rho=\ket{\Psi(t)}\bra{\Psi(t)}$ is separable. When $\rho$ represents a pure 
bipartite system, the R\'enyi entropies of its subsystems are equal, 
$S_\al(\rho_A)=S_\al(\rho_B)$. The R\'enyi entropies form a monotonically 
decreasing series in $\al$ since $\pa S_\al/\pa\al\leq 0$.

 In the remainder of this section, we derive an entanglement timescale for the 
R\'enyi entropies of a pure bipartite state \eqref{eqSt} evolving under a 
general Hamiltonian \eqref{eqHam}. Initially the subsystems are pure, $
\rho_A\ {=}\ \rho_A^2$, because \eqref{eqSt} is separable, and therefore 
$S_\al(\rho_A)|_{t=0}{=}\frac{1}{1-\al} \ln \tr_A \rho_A^\al |_{t=0} {=}0$. 
As the state evolves, the interactions between subsystems will generate 
entanglement. Starting at a minimum of $S_\al$, the first time derivative is 
initially zero. We will calculate the second derivative to obtain a Taylor 
expansion around $t=0$ of the form 
\begin{equation}
S_\al(\rho_A) =C_\al\frac{t^2}{T_{\mathrm{ent}}^2}+O(t^3).
\end{equation}
We will find that the entanglement timescale 
$T_{\mathrm{ent}}$ takes the same form for all R\'enyi entropies, with 
$C_\al$ a constant.

Since the R\'enyi entropies are initially minimal, their first derivatives must 
vanish. We find $\frac{d}{dt}S_\al(\rho_A){=}\frac{\al}{1-\al}
\left(\tr_A\rho_A^\al\right)^{-1}\tr_A\big[(\tr_B\rho)^{\al-1}{\tr_B}{\big(\parder{\rho
}{t}\big)}\big]$. Note that in general, $\left[{\tr_B}{\left(\pa\rho/\pa t\right)},
\tr_B\rho\right]\ {\neq}\ 0$. However, inside the $A$ trace, we can cyclically 
permute each term produced by the derivative into a common ordering as 
shown. Using the von Neumann equation $\pa\rho/\pa t =-i [H,\rho]$ with $
\hbar=1$ and using \eqref{eqHam} and \eqref{eqSt} in the $t=0$ limit, we find 
$\frac{d}{dt}S_\al(\rho_A)|_{t=0}=\frac{i\al}
{\al-1}{\left(\tr_A\rho_A^\al\right)^{-1}}{\sum_n}\tr_B(\rho_B B_n)
\tr_A\big(\rho_A^{\al-1}A_n\rho_A{-}\rho_A^\al A_n\big){=}0$.

\begin{widetext}
The leading order of the time evolution comes from the second derivative,
\begin{align}\label{eq2nd}
\begin{aligned}
\frac{d^2}{dt^2}S_\al(\rho_A)\ =\frac{1}{1-\al}
\bigg(\left(\tr_A\rho_A^\al\right)^{-1}\tr_A\bigg\{\frac{d^2}{dt^2}
[\tr_B\rho(t)]^\al\bigg\}-\left(\tr_A\rho_A^\al\right)^{-2}\bigg\{\tr_A\frac{d}{dt}
[\tr_B\rho(t)]^\al\bigg\}^2\bigg).
\end{aligned}
\end{align}
 The second term vanishes when the $t\to0$ limit is taken; this was the result 
of the first derivative calculation. We are left with the first term of 
\eqref{eq2nd} for which we find
\begin{align}\label{eq2}
\begin{aligned}
\tr_A\bigg\{\frac{d^2}{dt^2}
[\tr_B\rho(t)]^\al\bigg\}=\al\tr_A\left[(\tr_B\rho)^{\al-1}\tr_B\parder{^2\rho}{t^2}+
\sum_{\be=0}^{\al-2}(\tr_B\rho)^{\be}\tr_B\parder{\rho}{t}(\tr_B\rho)^{\al-2-\be}
\tr_B\parder{\rho}{t}\right].
\end{aligned}
\end{align}
 The $\be$ sum keeps track of the non-commuting factors which cannot be 
permuted into a common ordering. Applying the von Neumann equation 
leads to 
\begin{align}\label{eq3}
\begin{aligned}
\frac{d^2}{dt^2}S_\al(\rho_A)\big|_{t=0}\ &{=}\frac{\al}
{\al-1}\left(\tr_A\rho_A^\al\right)^{-1}\sum_{n,m}\bigg[\tr_B(B_nB_m\rho_B)
\tr_A\left(2A_nA_m\rho_A^\al-2A_m\rho_AA_n\rho_A^{\al-1}\right)\\
+&\tr_B(B_n\rho_B)\tr_B(B_m\rho_B){\sum_{\be=0}
^{\al-2}}{\tr_A}{\left(2\rho_A^{\be+1}A_n\rho_A^{\al-\be-1}A_m-\rho_A^\be 
A_n\rho_A^{\al-\be}A_m-\rho_A^{\be+2}A_n\rho_A^{\al-2-\be}A_m\right)}
\bigg].
\end{aligned}
\end{align}

Before simplifying \eqref{eq3} for general $\al$, it is useful to look at the 
unique case of $\al=2$ which corresponds to the purity studied in 
\cite{Yang2017a}. In this case, the $\be$ sum contains only a single term. 
Using the assumption of purity at $t=0$ allows us to write
\begin{align}\label{eq5}
\begin{aligned}
\frac{d^2}{dt^2}S_2(\rho_A)\big|_{t=0} =4\sum_{n,m}
\left[\tr_B(B_nB_m\rho_B)-\tr_B(B_n\rho_B)\tr_B(B_m\rho_B)\right]
\left[\tr_A(A_nA_m\rho_A)-\tr_A(A_m\rho_AA_n\rho_A)\right].
\end{aligned}
\end{align}
Note that we have \emph{not} assumed that $[A_n,A_m]=0$. Instead, we 
have used the symmetry of $\tr_B(B_n\rho_B)\tr_B(B_m\rho_B)$ in the $n$, $m$ indices to exchange $A_n$ and $A_m$. Indeed, \eqref{eq5} exactly 
matches the main result of \cite{Yang2017a} when we account for the 
difference in the definitions of the purity and R\'enyi entropy. Defining the $
\al$ purity, $P_\al(\rho_A)=\tr_A\rho_A^\al$, we have under our assumptions $
\frac{d^2}{dt^2}S_\al(\rho_A)|_{t=0}=\frac{1}{1-\al}\frac{d^2}{dt^2}
P_\al(\rho_A)|_{t=0}.$

Returning to the general case, it is possible to greatly simplify \eqref{eq3} by 
using the idempotency of $\rho_A(t=0)$, and $\rho_A^0=\id_A$ where $
\id_A$ is the identity operator for subsystem $A$. The special case of $
\rho_A^0=\id_A$ only occurs in the $\be$ sum when $\be$ takes on its 
extreme values of 0 and $\al{-}2$. Each other term in the sum vanishes. The 
general result for ${\al{>}2}$ is
\begin{align}\label{eq6}
\begin{aligned}
\frac{d^2}{dt^2}S_{\al}(\rho_A)\big|_{t=0} &=\frac{2\al}{\al-1}\sum_{n,m}
\left[\tr_B(B_nB_m\rho_B)-\tr_B(B_n\rho_B)\tr_B(B_m\rho_B)\right]
\left[\tr_A(A_nA_m\rho_A)-\tr_A(A_m\rho_AA_n\rho_A)\right]\\
&=\frac{2\al}{\al-1}\sum_{n,m}\left[\langle B_nB_m \rangle-\langle B_n 
\rangle\langle B_m \rangle\right]\left[\langle A_nA_m \rangle-\langle A_n 
\rangle\langle A_m \rangle\right]=\frac{2\al}{\al-1}T_{\mathrm{ent}}^{-2},
\end{aligned}
\end{align}
 where we have used the simplification $
\tr_A(A_m\rho_AA_n\rho_A)=\tr_A(A_m\rho_A)\tr_A(A_n\rho_A)$ for pure $
\rho_A$ as shown in \cite{Yang2017a}.

\end{widetext}

Equation \eqref{eq6} is our main result and shows that the second derivative 
of every R\'enyi entropy for $\al>2$ is of the same universal form as the $
\al=2$ case studied previously. In fact, the coefficient incorporates the $\al=2$ 
case in Eq. \eqref{eq5} as well. The only remaining case is $\al=1$, which we 
turn to now.

The entanglement entropy $S(\rho_A)=-\tr_A(\rho_A\ln\rho_A)$ is the most 
widely used entanglement measure in the literature. It corresponds to the $
\al\to1^+$ limit of $S_\al(\rho_A)$ after an analytic continuation in $\al$ 
\cite{Calabrese2004,Calabrese2009}. Inserting $\al=1$ at intermediate steps 
in the derivation leading to \eqref{eq6} produces ill-defined quantities since 
the density matrix $\rho_A(t=0)$ is pure, and therefore singular. 
Nevertheless, we emphasize that inverse powers of $\rho_A$ do not appear 
in the final result \eqref{eq6}. The prefactor $2\al/(\al-1)$ can be analytically 
continued in $\al$ and is analytic away from the simple pole at $\al=1$. 
Taking the limit of $2\al/(\al-1)$ as $\al\to1^+$ along the real axis shows that 
$d^2S(\rho_A)/dt^2|_{t=0}$ is proportional to the entanglement timescale 
with a divergent prefactor. This reflects the entanglement entropy's sensitivity 
to small eigenvalues of $\rho_A$ via the logarithm.

 To make this point more clear, let $p_i(t)$ be the eigenvalues of $\rho_A$ 
such that $p_1(0)=1$ and $p_j(0)=0$ ($j\neq 1$). Then the second derivative 
of the entanglement entropy, $S(\rho_A)=-\sum (p_i \ln p_i)$, in the $t\to 0$ 
limit is
\begin{equation}\label{eqEET}
\frac{d^2 S}{dt^2}=-\frac{d^2p_1}{dt^2}-\sum_{j\neq1}\left[( \ln 
p_j+3)\frac{d^2p_j}{dt^2}\right].
\end{equation}
 Generically, $\lim_{t\to0}(d^2p_j/dt^2) \ln p_j$ is divergent since $d^2p_j/
dt^2$ is not required to be zero initially. Still, the divergence of $d^2 S/dt^2$ 
at $t=0$ does not imply that the entanglement entropy itself diverges; on the 
contrary, $S(\rho_A)$ is strictly bounded above by the dimension of the 
Hilbert space of subsystem $A$. Rather, $d^2 S/dt^2$ appears in the Taylor 
series as the coefficient of $t^2$ which tames the logarithmic divergence. It 
should be noted that higher derivatives also diverge logarithmically at $t=0$, 
but are suppressed by higher powers of $t$.

\section{Example - Jaynes-Cummings model}

Equation \eqref{eqEET} shows that the divergence of $d^2S/dt^2$ at $t=0$ 
for an initially pure product state found in \eqref{eq6} is not an artifact of the 
analytic continuation in $\al$. This is the generic behavior of the 
entanglement entropy for an initially separable state. To explore the physical 
significance of the entanglement timescale, and to check the divergence of 
$d^2S/dt^2|_{t=0}$, we work with the Jaynes-Cummings model (JCM) of a 
two-level atom interacting with a quantized radiation field 
\cite{Jaynes1963,Shore1993}. This system has been extensively studied in 
quantum optics because of its interesting entanglement properties 
\cite{Phoenix1988,Quesada2013} and quantum revivals 
\cite{Eberly1980,Pimenta2016}. In this section, we calculate the entanglement 
timescale for initially separable states, first by finding an analytic solution for 
the R\'enyi entropies at all times, and then by studying the expectation values 
of the interaction terms in the initial state as dictated by \eqref{eq6}. We 
explicitly show that the divergence of $d^2S/dt^2|_{t=0}$ is only logarithmic.

In the rotating-wave approximation, the JCM Hamiltonian is \cite{Shore1993}
\begin{equation}
\frac{H}{\hbar}=\frac{\om_0}{2}\si_z+\om a^\dag a+\la(a^\dag \si_-+a\si_+).
\end{equation}
\n Here, $\om_0$ is the atomic transition frequency, $\om$ is the characteristic 
field frequency, and $\la$ is a coupling constant. For simplicity, we impose the 
resonance condition $\om=\om_0$ and set $\hbar=1$. The Pauli operators 
can be written in terms of the atomic ground state $\ket{g}$ and excited state 
$\ket{e}$ as $\si_z=\ket{e}\bra{e}-\ket{g}\bra{g}$, $\si_-=\ket{g}\bra{e}$, and $
\si_+=\ket{e}\bra{g}$. The field mode has a Fock basis $\ket{n}$ on which the 
creation and annihilation operators $a^\dag$, $a$ act in the usual way. Notice 
that this Hamiltonian is of the assumed product form \eqref{eqHam} and is 
time independent.

\begin{figure*}
    \centering
    \begin{subfigure}[b]{0.49\textwidth}
        \includegraphics[width=0.8\textwidth]{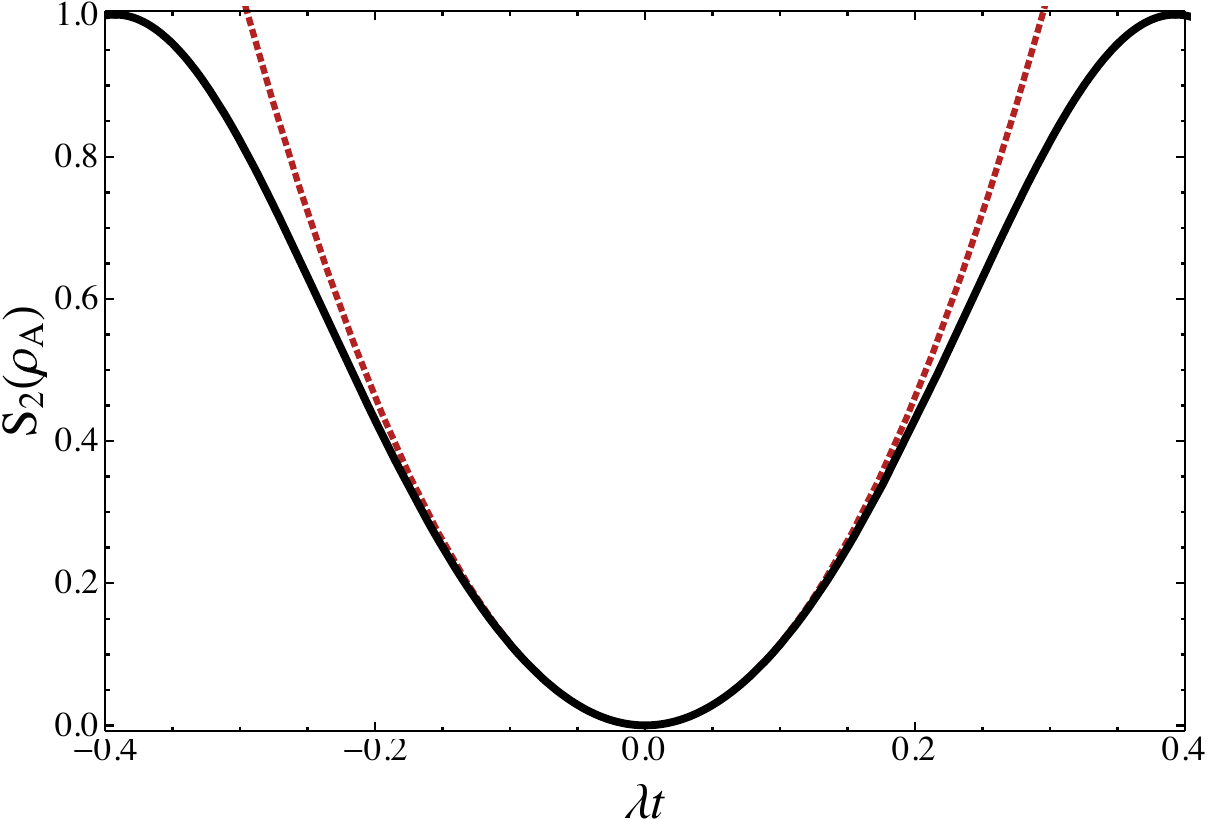}
        \label{fig:focka}
    \end{subfigure}
    \begin{subfigure}[b]{0.49\textwidth}
        \includegraphics[width=0.8\textwidth]{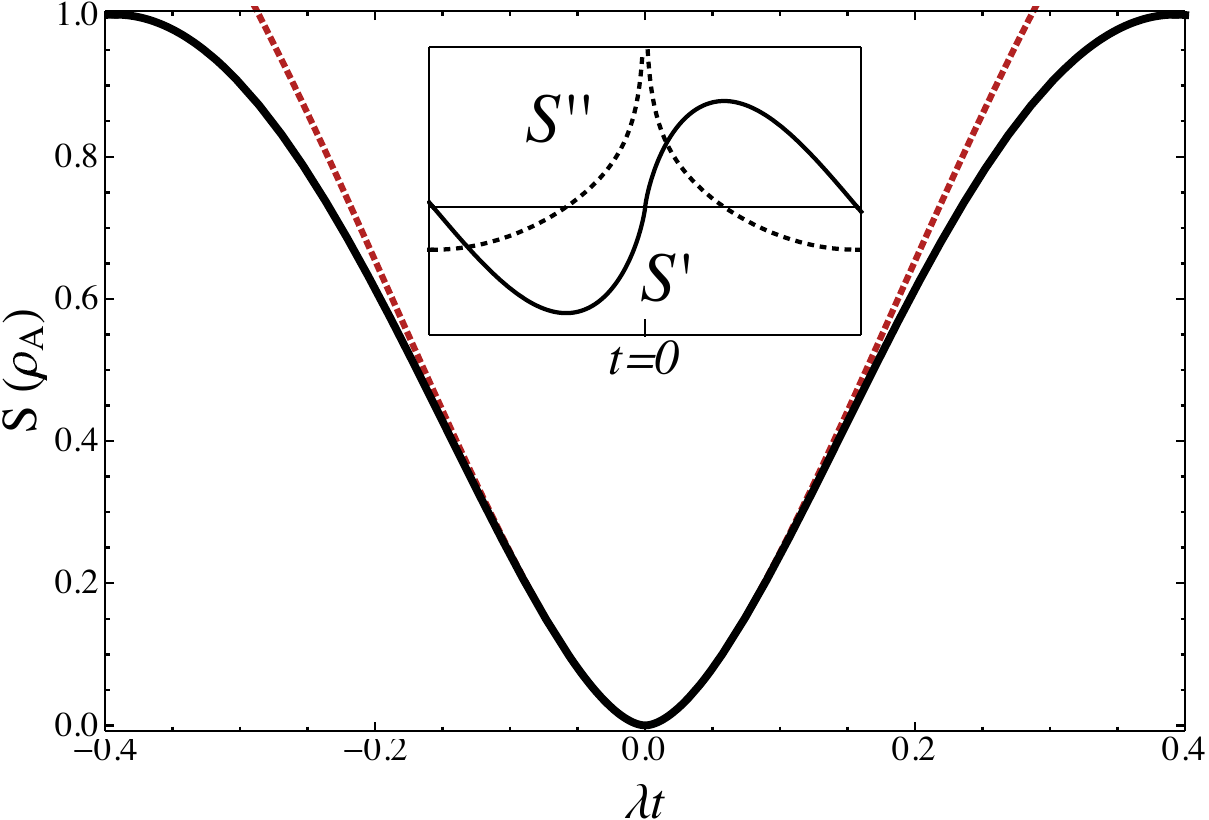}
         \label{fig:fockb}
    \end{subfigure}
    \caption{
   \small (a) $S_2(\rho_A)$ for the Fock state with $N=3$ and 
        $C_e=1$ is sinusoidal and $C^\infty$ smooth. $S_2$ is compared to 
        the quadratic approximation with 
        timescale $\la T_{\mathrm{ent},e}=1/4$ (dashed red line). (b) $S(\rho_A)$ 
        for the same state is differentiable, but $d^2 S/dt^2$ is discontinuous 
        at $t=0$ (inset, dashed line). Units of $\ln(2)$ are used in all figures. 
         \normalsize	
        }
        \label{fig:fock}
\end{figure*}

Let the overall initial state be the product of an arbitrary atomic state 
$\ket{\psi}_A=C_g\ket{g}+C_e\ket{e}$ and field state $\ket{\psi}_F{=}\sum_{n=0}
^\infty C_n \ket{n}$. Then the overall state at any time is 
\cite{Gea-Banacloche1990}
\small
\begin{align}\begin{aligned}\label{eqjcmstate}
&\ket{\Psi(t)}\\
&={\sum_{n=0}^\infty}\{[C_e{C_n} \cos( {\la} \sqrt{n+1} t 
) {-}iC_g C_{n+1}\sin( {\la}\sqrt{n+1} t )] \ket{e}\\
 &\ \ {+}[ -iC_eC_{n-1}\sin( {\la}\sqrt{n}t)+C_g C_n\cos( {\la}\sqrt{n}t 
)]\ket{g}\}\ket{n},
\end{aligned}\raisetag{\baselineskip}\end{align}
\normalsize
 which is entangled for most times. Since the exact solution for the state is 
available, the R\'enyi entropies can be calculated directly for either 
subsystem after a partial trace. When the atom is initially excited $(C_e{=}\ 1,\ 
C_g{=}\ 0)$,
\small
\begin{align}\label{eqjcme}\begin{aligned}
\frac{d^2}{dt^2}S_\al(\rho_A)\bigg|_{t=0}&=\frac{2\al}{\al-1} {\lambda 
^2}{\left[ \sum _{n=0}^{\infty } (n+1) \left| C_n\right| ^2\right.}\\
&\ - \left. {\sum _{n,m=0}^{\infty}} \sqrt{m+1} \sqrt{n+1} C_{n+1}^* C_{n} C_{m+1}C_{m}^*\right].
\end{aligned}\end{align}
\normalsize
 For comparison, if the atom is initially in the ground state, then the result in 
\eqref{eqjcme} changes slightly by the replacement $\left| C_n\right| ^2\to\left| 
C_{n+1}\right| ^2$ in the first sum.

The entanglement timescale can alternatively be computed from the 
Hamiltonian and initial state by using the definition in \eqref{eq6}. This is 
much simpler because it does not require solving for the time evolution of the 
system. When the atom is initially excited, the only nonzero term in 
\eqref{eq6} is
\small
\begin{align}\label{eqTe}\begin{aligned}
T_{\mathrm{ent},e}^{-2}&=\la^2(\langle a a^\dag  \rangle-\langle a 
\rangle\langle a^\dag \rangle)(\langle \si_+ \si_- \rangle-\langle \si_+ 
\rangle\langle \si_-\rangle)\\
&=\lambda ^2\left[ \sum _{n=0}^{\infty } (n+1) \left| C_n\right| ^2\right.\\
&\ \ \left. - \sum 
_{n,m=0}^{\infty } \sqrt{m+1} \sqrt{n+1} C_{n+1}^* C_{n} C_{m+1}C_{m}
^*\right]\geq1.
\end{aligned}\end{align}
\normalsize
 Similarly for the ground-state case, we find a single nonzero term,
\begin{equation}
T_{\mathrm{ent},g}^{-2}=\la^2(\langle a^\dag a\rangle-\langle a^\dag 
\rangle\langle a \rangle)(\langle \si_- \si_+ \rangle-\langle \si_- 
\rangle\langle \si_+\rangle)\geq0,
\end{equation}
which~is~like~{\eqref{eqTe}}~but~with~${| C_n|^2}{\to}{| C_{n{+}1}| 
^2}$~in~the~first~sum.

The growth of entanglement is always controlled by the strength of the 
coupling $\la$ between subsystems. Indeed, it was pointed out in early 
studies of the JCM that $\la^{-1}$ is proportional to the time period over 
which the reduced states remain approximately pure 
\cite{Gea-Banacloche1990}. The positivity of R\'enyi entropies requires that 
$T_{\mathrm{ent}}^{-2}$ is positive. This is ensured by the results of 
\cite{Yang2017a}, but can be seen here as a consequence of the 
Cauchy-Schwarz inequality which implies 
$\langle a^\dag a\rangle\geq\langle a^\dag \rangle\langle a \rangle$, etc.

From these general expressions, we can easily examine the growth of 
entanglement for some common field states. Consider when the field is 
initially in a Fock state, $\ket{\psi}_F=\ket{N}$. For the initially excited state, we 
find $T_{\mathrm{ent},e}=({\la\sqrt{N+1}})^{-1}$ and for the ground state, 
$T_{\mathrm{ent},g}=({\la\sqrt{N}})^{-1}$. Figure \ref{fig:fock} shows 
$S_2(\rho_A)$ and $S(\rho_A)$ for $C_e=1,\ C_g=0$, and $N=3$, 
along with the quadratic timescale approximation. Whereas $S_\al(\rho_A)$ for 
$\al\geq2$ is $C^\infty$ smooth in this example, we see that 
$d^2 S(\rho_A)/dt^2$ diverges at $t=0$ as expected, while $dS(\rho_A)/dt$ 
is continuous at $t=0$.

\begin{figure*}
    \centering
    \begin{subfigure}[b]{0.49\textwidth}
        \includegraphics[width=0.8\textwidth]{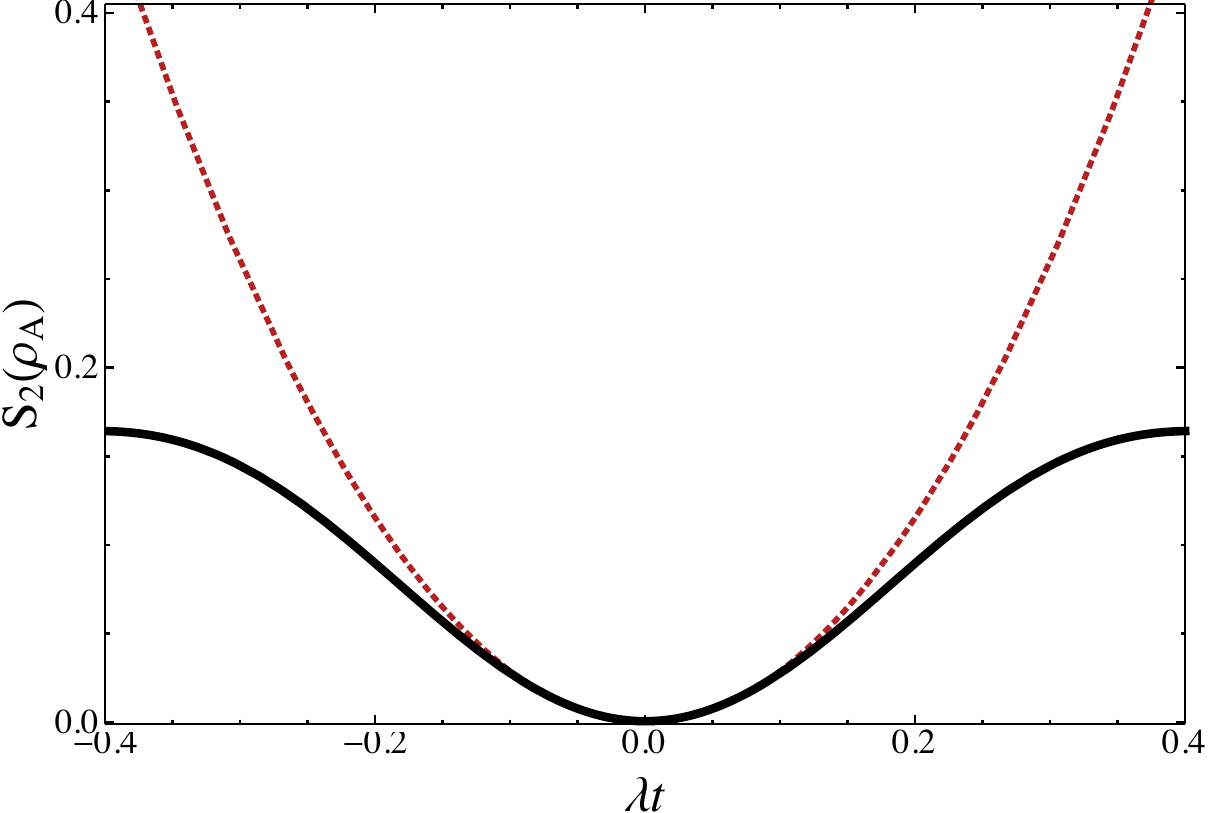}
         \label{fig:cohEa}
    \end{subfigure}
    \begin{subfigure}[b]{0.49\textwidth}
        \includegraphics[width=0.8\textwidth]{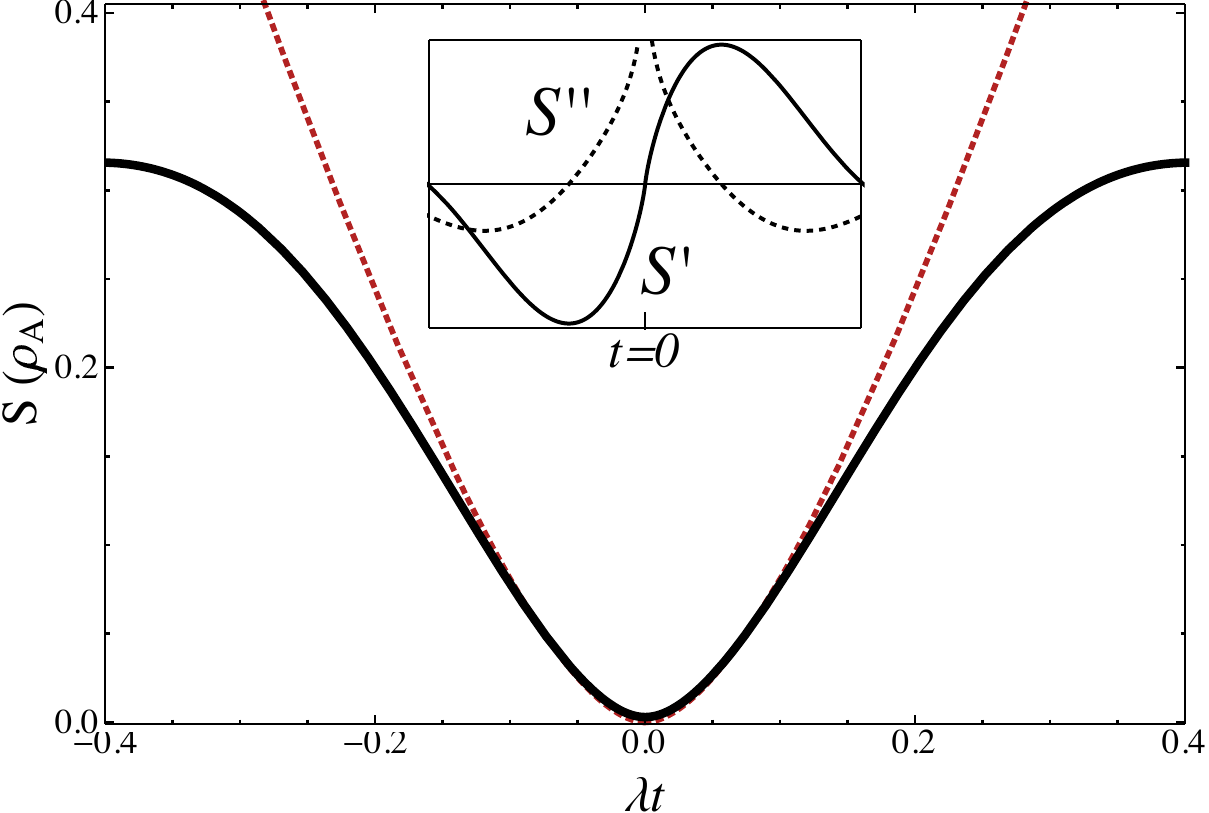}
         \label{fig:cohEb}
    \end{subfigure}
    \caption{\small (a) $S_2(\rho_A)$ for the coherent state with $\nu=3$ and 
        $C_e=1$. The small-$t$ behavior is independent of $\nu$ and described 
        by the quadratic timescale $\la T_{\mathrm{ent},e}=1$ (dashed red line). 
        (b) $S(\rho_A)$ 
        for the same state is differentiable, but $d^2 S/dt^2$ is discontinuous 
        at $t=0$ (inset dashed line).
    \normalsize}\label{fig:cohE}
\end{figure*}

 Instead, if the field starts in a coherent state,
\begin{equation}
\ket{\psi}_F=e^{-\tfrac{1}{2}|\nu|^2}\sum_{n=0}^\infty \frac{\nu^n}{\sqrt{n!}}
\ket{n},\quad a\ket{\psi}_F=\nu\ket{\psi}_F,
\end{equation}
 then the excited state timescale is $T_{\mathrm{ent},e}{=}\ {1/\la}$, whereas for 
the ground state, $T_{\mathrm{ent},g}^{-1}{=}\ 0$. Notably, these timescales are 
independent of $\nu$. Figure \ref{fig:cohE} shows $S_2(\rho_A)$ and 
$S(\rho_A)$ for the coherent state with $\nu\ {=} \ 3$ and $C_e\ {=}\ 1$,~$C_g\ {=}\ 0$. 
Once again, $d^2 S(\rho_A)/dt^2$ diverges at $t\ {=}\ 0$, while $dS(\rho_A)/dt$ is continuous at $t=0$. 

\begin{figure*}
    \centering
    \begin{subfigure}[b]{0.49\textwidth}
        \includegraphics[width=0.8\textwidth]{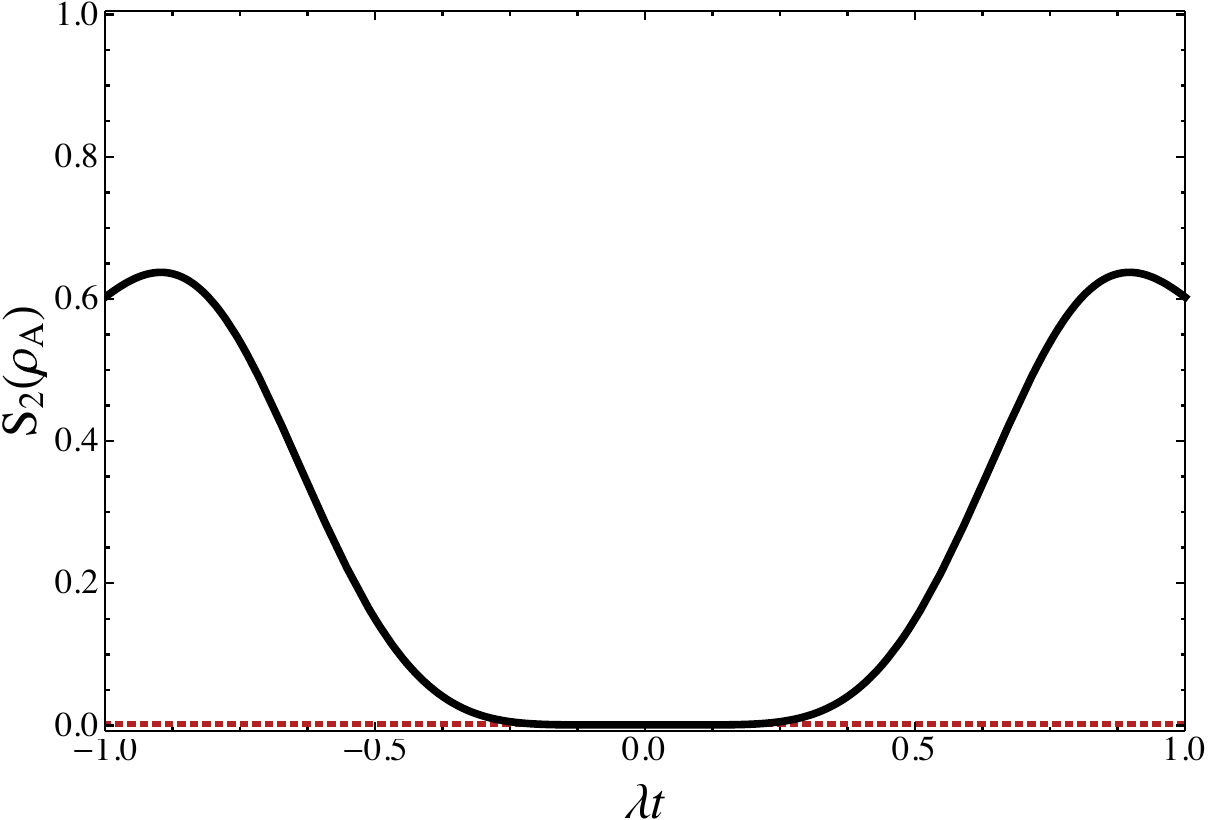}
         \label{fig:cohGa}
    \end{subfigure}
    \begin{subfigure}[b]{0.49\textwidth}
        \includegraphics[width=0.8\textwidth]{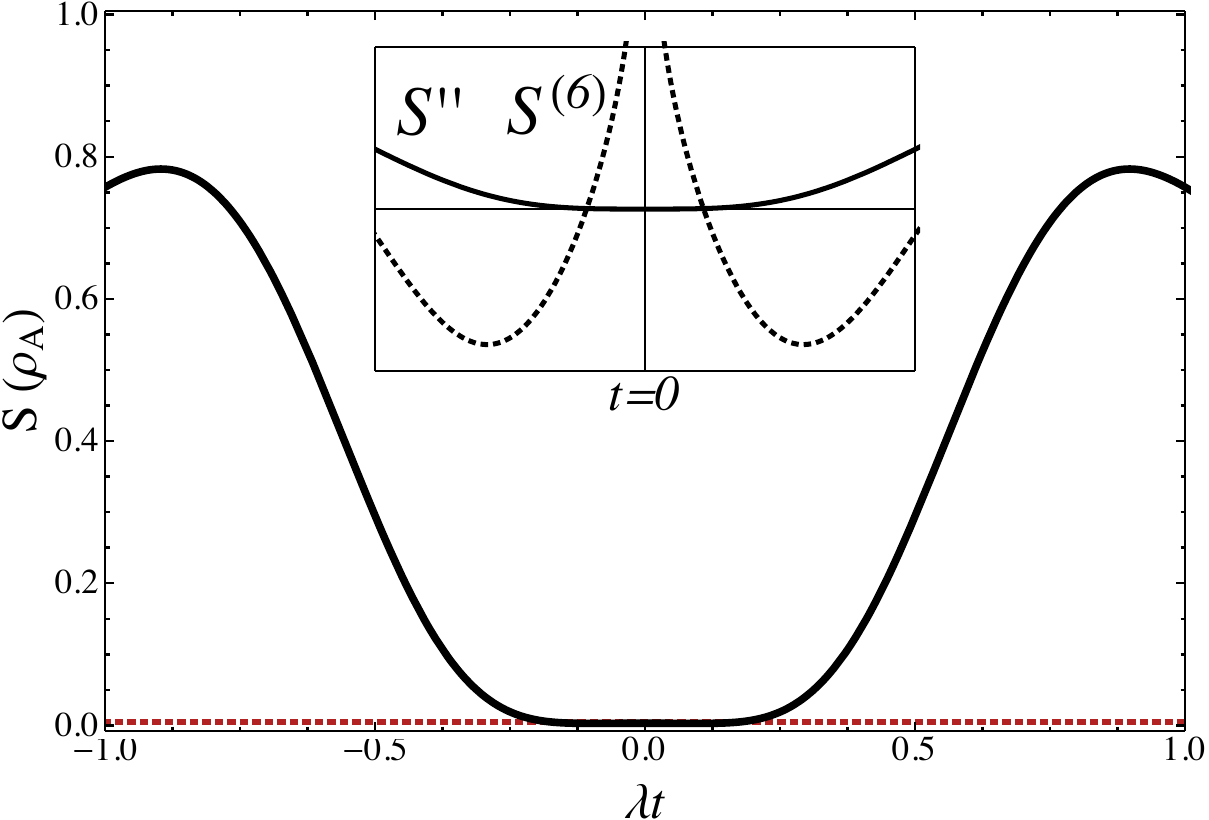}
       \label{fig:cohGb}
    \end{subfigure}
     \caption{\small (a) $S_2(\rho_A)$ for the coherent state with $\nu=3$ and 
        $C_g=1$, where $T_{\mathrm{ent},g}^{-1}=0$
        indicates that the state remains effectively separable for a significant time. 
        The leading behavior 
        around $t=0$ is sixth order in $t$. (b) $S(\rho_A)$ 
        for the same state is $C^5$ smooth, with $d^2 S/dt^2|_{t=0}=0$ 
        (inset, solid line), and $d^6 S/dt^6$ discontinuous 
        at $t=0$ (inset, dashed line).\normalsize}\label{fig:cohG}
\end{figure*}

For comparison, the coherent state with $\nu=3$ and $C_e=0$, $C_g=1$ 
remains effectively separable for some time, as shown in 
Fig. \ref{fig:cohG}. The divergence of the 
entanglement timescale in this case means one must look to higher orders in 
the Taylor expansion of $S_\al(t)$ to see the growth of entanglement. This is 
one example of an initial state where the correlated quantum uncertainty 
defined in \cite{Yang2017a} vanishes. 

Equation \eqref{eqjcme} shows that the second time derivative of the 
entanglement entropy typically will be divergent in separable states. 
This is not a flaw of  taking the $\al\to1$ limit of the R\'enyi entropy, 
but is the actual behavior of the 
entanglement entropy. From the state \eqref{eqjcmstate}, we can calculate 
the entanglement entropy directly for all times by diagonalizing the reduced 
density matrix of the atom $\rho_A(t)$ and finding its eigenvalues,
$p_{1}(t)=\frac{1}{2}(1+|\vec s(t)|)$, and $p_{2}(t)=\frac{1}{2}(1-|\vec s(t)|)$ in 
terms of the Bloch vector $\vec s(t)$ \cite{Gerry2005}. For instance, starting 
with the atom in its excited state, we find $d^2p_1/dt^2|_{t=0}
=-2T_{\mathrm{ent},e}^{-2}=-d^2p_2/dt^2|_{t=0}.$ Using \eqref{eqEET} leads to the logarithmically divergent result,
\small
\begin{align}\begin{aligned}
& \frac{d^2 S}{dt^2}\bigg|_{t=0}\\
 &\ {=}2\left\{{-2{+}\ln 2{-}\lim_{t\to0}{\ln}{\left[1{-}
{\sum_{n=0}^\infty} |C_{n}|^2\cos^2(\la \sqrt{n+1}  t)\right]}}\right\}T_{\mathrm{ent},e}^{-2}.
\end{aligned}\end{align}
\normalsize
A similar logarithmic divergence occurs for the atom initially in its ground 
state. 

\section{Discussion}\label{dis}

The main result of \cite{Yang2017a} showed that for any unentangled pure 
bipartite state evolving under an arbitrary Hamiltonian, the growth of 
entanglement is characterized by a timescale which takes the universal form
 \small
\begin{equation}
T_{\mathrm{ent}}{=}
\left[\sum_{n,m}\left(\langle A_nA_m \rangle{-}\langle A_n \rangle\langle A_m 
\rangle\right)\left(\langle B_nB_m \rangle{-}\langle B_n \rangle\langle B_m 
\rangle\right)\right]^{-\tfrac{1}{2}}
\end{equation}
\normalsize
 where entanglement is measured by the purity of subsystems. In this paper, 
we have shown that the same timescale characterizes the growth of 
entanglement as measured by any R\'enyi entropy. Since the family of R\'enyi 
entropies constitutes a complete determination of the entanglement in a pure 
bipartite system, the entanglement timescale universally describes the initial 
growth of bipartite entanglement.

It is easy to prove that the entanglement timescale obeys several properties 
expected of the R\'enyi entropy. As shown in \cite{Yang2017a}, 
$T_{\mathrm{ent}}^2$ is a manifestly positive quantity so that the R\'enyi 
entropies initially increase from their minimum value. It is also symmetric 
between the subsystems $A$ and $B$ which reflects the symmetry 
$S_\al(\rho_A)=S_\al(\rho_B)$ for overall pure states. Furthermore, the 
coefficient ${2\al}/(\al-1)$ in \eqref{eq6} is monotonically decreasing in $\al$, 
which is required by the general condition $\pa S_\al/\pa \al\leq 0$.

R\'enyi entropies are widely used theoretically and have recently been 
measured in isolated many-body systems \cite{Islam2015a}, including their 
time dependence after an interaction is turned on \cite{Kaufman2016}. The 
first such measurement was performed on a Bose-Einstein condensate 
trapped in an optical lattice and evolving under the Bose-Hubbard 
Hamiltonian in one dimension,
\begin{equation}\label{eqBH}
H=-J\sum_{\langle i,j \rangle} a^\dag_i a_j +\frac{U}{2}\sum_{i} a^\dag_i 
a_i(a^\dag_i a_i-1).
\end{equation}
 The first sum is over nearest-neighbor pairs and represents tunneling 
between neighboring sites at a rate $J$. The second sum over each lattice 
site represents the attractive energy among bosons sharing a site. In the 
experiment \cite{Kaufman2016}, a product of one-particle Fock states was 
prepared on six adjacent lattice sites with a barrier on each end. After a 
quench in which the interaction in \eqref{eqBH} was turned on, the second 
R\'enyi $S_2(\rho_A)$ was measured in time for all unique partitions of the 
six sites.

The only interaction term in \eqref{eqBH} that couples $A$ to $B$ is $-
J(a^\dag_i a_{i+1}+a_i a_{i+1}^\dag)$, where sites $i$ and $i {+} 1$ are 
neighbors across the partition. Thus, for any nontrivial partitioning, the 
entanglement timescale is the same, $T_{\mathrm{ent},BH}^{-2}
={J^2}{\braket{1|a_i^\dag a_i|1}}{\braket{1|a_{i{+}1} a_{i{+}1}^\dag|1}}~{+}~{J^2}{\braket{1|a_i a_i^\dag|1}}{\braket{1|a_{i{+}1}^\dag a_{i{+}1}|1}}{=}$ $4 J^2$. Using the 
experimental value of $J/2\pi=66$ Hz, we can estimate that the entanglement 
will become significant within a time $T_{\mathrm{ent},BH}=1.2\ \mathrm{ms}
$, which agrees with the experimental result displayed in Fig. 3 of Ref. 
\cite{Kaufman2016}. This comparison is only approximate since the actual 
initial states prepared in the experiment were not free of entanglement.

The original motivation to determine the entanglement timescale was to 
estimate how quickly a generic quantum system will decohere due to 
entanglement with gravitational degrees of freedom 
\cite{Zurek2003,Baker2017,Yang2017}. This question is relevant to the black-hole information problem \cite{Almheiri2013,Harlow2016}, where the 
Hawking quanta escaping from the black-hole horizon region may entangle 
with the geometry itself. To make any concrete statements about 
entanglement with gravitational degrees of freedom, one needs to work with 
quantum field theory or, better yet, quantum gravity. Since our derivation of 
the entanglement timescale assumes that the initial state is pure and 
unentangled, it is difficult to generalize these results to quantum field theory, 
where typical states are highly entangled on all scales 
\cite{Bombelli1986,Srednicki1993,Holzhey1994}. UV divergent 
entanglements can be avoided by considering the entanglement difference 
between states, for example with the relative entropy, which lends hope for 
our analysis of $d^2S_\al/dt^2$ \cite{Lashkari2014,Ruggiero2017}. One can 
otherwise avoid divergences by considering causally separated subregions, 
but this comes at the cost of losing purity for the combined system 
\cite{Hollands2017}. Moreover, for gauge field theories, the Hilbert space 
does not factorize across spatial boundaries, invalidating our assumptions 
\cite{Casini2014,Donnelly2014}. Still, the growth of entanglement in quantum field theory 
states is a major area of research in many-body, condensed-matter, and high-energy physics \cite{Avery2014,Liu2014,Belin2013,Caputa2014}, and it 
would be interesting to develop an entanglement timescale in these regimes.

\begin{acknowledgments}
The author would like to thank I-Sheng Yang, Hudson Pimenta, Aaron 
Goldberg, and A.W. Peet for helpful discussions. The author is financially supported by 
a Vanier Canada Graduate Scholarship, an Ontario Graduate Scholarship, 
and a Discovery Grant from the Natural Sciences and Engineering Research 
Council of Canada.
\end{acknowledgments}

\bibliography{Renyi_PRA.bib}

\end{document}